\documentclass[conference]{IEEEtran}
\IEEEoverridecommandlockouts

\usepackage[utf8]{inputenc}
\usepackage[T1]{fontenc}
\usepackage{cite}
\usepackage{amsmath,amssymb,amsfonts}
\usepackage{graphicx}
\usepackage{booktabs}
\usepackage{multirow}
\usepackage{array}
\usepackage{tabularx}
\usepackage{makecell}
\usepackage[caption=false,font=footnotesize]{subfig}
\usepackage{xcolor}
\usepackage{url}

\ifdefined\XeTeXversion
\else
  \DeclareUnicodeCharacter{00B1}{\ensuremath{\pm}}
  \DeclareUnicodeCharacter{2265}{\ensuremath{\geq}}
  \DeclareUnicodeCharacter{2264}{\ensuremath{\leq}}
  \DeclareUnicodeCharacter{2013}{--}
  \DeclareUnicodeCharacter{2014}{---}
  \DeclareUnicodeCharacter{2192}{\ensuremath{\rightarrow}}
\fi

\graphicspath{{figures/}}

\title{Evaluating Generalization and Long-Term Stability of Smartwatch-Based Cuffless Blood Pressure Monitoring}

\author{
\IEEEauthorblockN{
Hongda Huang\textsuperscript{1$\dagger$}, Xiao Peng\textsuperscript{1$\dagger$}, Xiaoyu Li\textsuperscript{2},
Shuailong Tang\textsuperscript{1}, Guangpu Zhu\textsuperscript{2},\\
Xiajiao Yang\textsuperscript{3}, Hongwei Li\textsuperscript{3}, Yelei Li\textsuperscript{2},
and Yali Zheng\textsuperscript{1*}}
\IEEEauthorblockA{
\textsuperscript{1}College of Health Science and Environmental Engineering, Shenzhen Technology University, Shenzhen, China\\
\textsuperscript{2}OPPO Health Lab, Shenzhen, China\\
\textsuperscript{3}Tibet Branch Hospital of West China Hospital, Sichuan University, Chengdu, China\\
\textsuperscript{$\dagger$}Equal contribution. \textsuperscript{*}Corresponding author: Yali Zheng (\url{zhengyali@sztu.edu.cn})}
}

\begin{document}

\maketitle

\begin{abstract}
Smartwatch-based cuffless blood pressure (BP) monitoring is a promising low-burden solution for long-term cardiovascular management, but practical deployment remains constrained by model complexity, uncertain cross-dataset generalization, and insufficient evidence on long-term tracking stability. This study presents an efficient Signal-Context Gated Fusion Network (SCGBPNet) for cuffless BP estimation from photoplethysmography (PPG) signals. SCGBPNet integrates signal-level representations through self-attention and further incorporates subject-specific contextual information via a compact gated fusion module. On Aurora-BP, SCGBPNet achieved SBP/DBP mean absolute errors (MAEs) of 7.08/4.90 mmHg, compared with 6.44/4.89 mmHg under within-dataset validation on the private smartwatch dataset, indicating a modest within-dataset discrepancy. Cross-dataset transfer from Aurora-BP to the smartwatch dataset increased the MAEs to 8.98/6.47 mmHg, demonstrating substantial cross-dataset degradation. For long-term stability, the proposed method showed low week-to-week performance variation and a low daily decay rate, indicating good stability over a follow-up period of more than one month. Nevertheless, estimation accuracy deteriorated when BP deviated substantially from an individual's baseline. These findings indicate that favorable aggregate MAE alone should not be interpreted as evidence of deployment-ready performance. More systematic and stratified evaluation across dataset settings, BP ranges, and baseline-deviation levels is needed to assess the practical reliability of smartwatch-based cuffless BP monitoring.
\end{abstract}


\section{Introduction}
\IEEEPARstart{H}{ypertension} remains a major risk factor for cardiovascular disease, yet a substantial proportion of adults with hypertension remain unaware of their condition \cite{ref1}. Current blood pressure (BP) assessment still relies largely on repeated office BP measurements \cite{ref2} or 24-hour ambulatory BP monitoring \cite{ref3}. Although these cuff-based methods are clinically established, they provide intermittent measurements and may interfere with daily activities and sleep, limiting their suitability for low-burden long-term BP tracking \cite{ref4}.

Wearable cuffless BP monitoring has therefore attracted increasing attention. Physiological signals such as photoplethysmography (PPG), electrocardiography (ECG), and pulse pressure waveforms have been used as surrogate measurements for BP estimation \cite{ref5}. Among them, PPG is particularly attractive for consumer wearables because of its low cost, simple optical sensing, and ease of integration. With the emergence of large-scale wearable datasets such as Aurora-BP and CAS-BP \cite{ref6,ref7}, data-driven methods have become the mainstream approaches for modeling the nonlinear relationship between physiological waveforms and BP \cite{ref8}.

Existing data-driven methods include convolutional neural networks (CNNs) for local waveform pattern extraction \cite{ref9,ref10}, recurrent models for temporal dependency modeling \cite{ref11,ref12}, and Transformer-based architectures for long-range representation learning \cite{ref13,ref14,ref15}. In addition to waveform modeling, recent studies have fused handcrafted PPG features, deep representations, and demographic variables to improve BP estimation accuracy \cite{ref15}. However, increasingly complex backbones and fusion strategies may introduce substantial computational cost, which is undesirable for resource-constrained wearable deployment. 

Beyond modeling strategies, long-term tracking capability and calibration behavior are critical for long-term use because most cuffless BP models are calibration dependent and their accuracy may change over time. Although a recent large-scale smartwatch study demonstrated the feasibility of cuffless BP tracking in real-world settings \cite{ref16}, it did not collect daily paired cuff reference BP measurements for accuracy validation, leaving time-dependent error drift, baseline-only calibration validity, and recalibration benefit insufficiently quantified.

Motivated by these limitations, this study proposes SCGBPNet, an efficient signal-context gated fusion network for smartwatch-based cuffless BP monitoring. Its generalization and long-term tracking stability are evaluated using two datasets collected under distinct experimental conditions: the public Aurora-BP dataset acquired under an in-office protocol and a private smartwatch dataset collected during real-world home BP monitoring for more than 30 days. The main contributions of this study are summarized as follows:
\begin{enumerate}
  \item We design SCGBPNet, a compact signal-context gated fusion architecture that efficiently integrates handcrafted features, deep waveform representations and subject-context information for smartwatch-based cuffless BP monitoring.
  \item We evaluate SCGBPNet on Aurora-BP and a private long-term smartwatch dataset using within-dataset validation and cross-dataset transfer testing to characterize generalization across experimental settings.
  \item We quantify long-term tracking behavior under baseline-only calibration and multiple recalibration intervals using week-to-week variation, daily performance-decay analysis, and baseline-deviation stratification, providing insights into individual recalibration frequency and the need to complement overall accuracy metrics with stratified analysis.
\end{enumerate}

\section{Methodology}
\subsection{Datasets and Preprocessing}
\begin{figure*}[!t]
  \centering
  \includegraphics[width=0.80\textwidth]{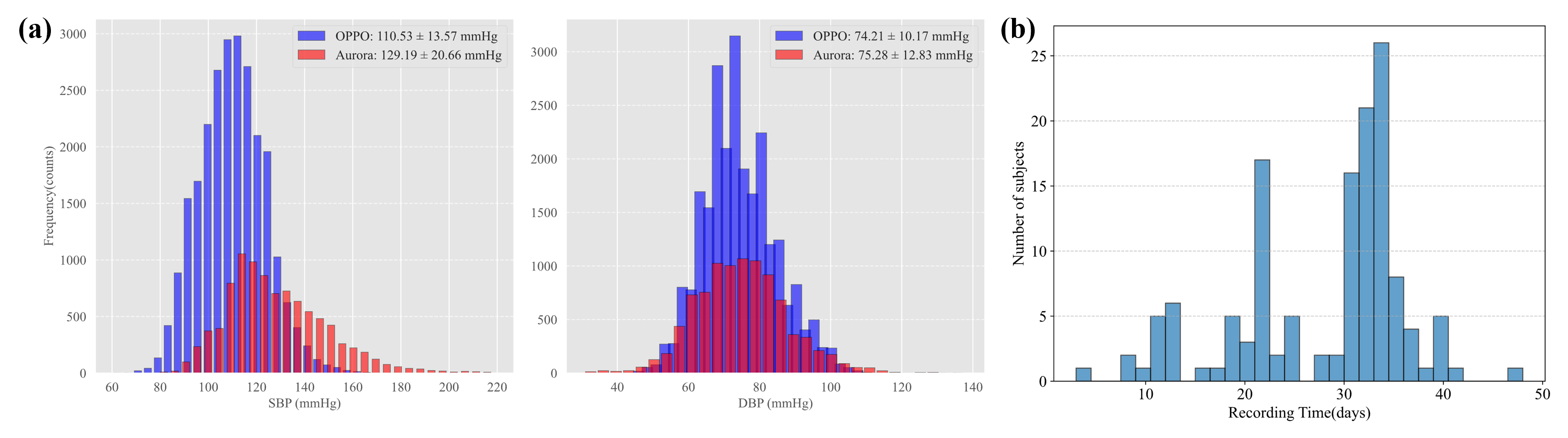}
  \caption{Dataset characteristics for BP distribution and longitudinal recording duration. (a) BP distributions of the two datasets; (b) distribution of recording duration per subject in the private smartwatch dataset.}
  \label{fig:dataset_characteristics}
\end{figure*}

As summarized in Table~\ref{tab:dataset_statistics} and Fig.~\ref{fig:dataset_characteristics}, Aurora-BP \cite{ref6} contained a broader BP distribution and an older population, whereas the private smartwatch dataset contained substantially more longitudinal recordings per subject. Table~\ref{tab:dataset_statistics} compares the two datasets with selected IEEE and ANSI/AAMI/ISO validation requirements \cite{ref24,ref25}. The details of the two datasets are described below.

\subsubsection{Aurora-BP Dataset}
The Microsoft Aurora-BP dataset includes in-clinic and ambulatory wearable-sensor measurements \cite{ref6}. To ensure consistency with the private smartwatch dataset, this study retained only the in-clinic subset, which includes static and post-exercise challenge states designed to induce BP changes. Reference BP values were obtained using an auscultatory device. Recordings were excluded if they contained missing entries ($n=427$), had a signal quality score below 0.5 ($n=1053$), or showed physiologically implausible BP values ($n=94$). After preprocessing and quality control, 9433 valid recordings from 626 participants were included in the analysis.

\subsubsection{Private Long-Term Smartwatch Dataset}
The private long-term smartwatch dataset was collected under a standardized home BP monitoring protocol approved by the Ethics Committee of the Tibet Branch of West China Hospital, Sichuan University (Research Ethics Approval No. 41, 2022). Written informed consent was obtained from all participants.

Single-channel wrist infrared PPG signals were acquired using OPPO smartwatches at 250 Hz, while reference BP was measured with an Omron electronic sphygmomanometer. Participants remained still during each measurement, and each valid reference BP measurement was paired with a 60-second PPG recording. Before each recording, participants rested for 5 min and were instructed to avoid intense exercise, caffeine intake, and measurements immediately after meals. Because monitoring took place at home over an extended period, these behavioral factors were not strictly controlled.

Participants performed at least three BP measurements during each of three daily periods (morning, noon, and evening), with at least 1 min between successive measurements. Additional readings were taken when consecutive measurements differed by more than 12 mmHg for SBP or 8 mmHg for DBP. Data collection lasted from several days to approximately 40 days, depending on the participant.

\begin{table*}[!t]
\centering
\caption{Dataset characteristics against standard requirements}
\label{tab:dataset_statistics}
\renewcommand{\arraystretch}{1.0}
\scriptsize
\setlength{\tabcolsep}{4pt}
\begin{tabularx}{\textwidth}{>{\raggedright\arraybackslash}m{0.17\textwidth} >{\centering\arraybackslash}m{0.12\textwidth} >{\centering\arraybackslash}m{0.14\textwidth} >{\raggedright\arraybackslash}m{0.24\textwidth} >{\raggedright\arraybackslash}m{0.24\textwidth}}
\toprule
Characteristic & Aurora-BP & Private smartwatch & \makecell[l]{IEEE 1708a-2019\\(legacy)} & \makecell[l]{ANSI/AAMI/ISO\\81060-2:2019} \\
\midrule
Number of participants & 626 & 136 & $\geq 85$ & $\geq 85$ \\
Number of BP recordings & 9433 & 24910 & $\geq 255$ pairs & $\geq 255$ pairs; $\geq 3$/subject \\
Age, years & $44.98 \pm 12.88$ & $33.38 \pm 6.52$ & Intended-use population & $>12$ years \\
Male/Female & 312/314 & 93/43 & $\geq 30\%$ each sex & $\geq 30\%$ each sex \\
SBP, mmHg & $129.19 \pm 20.66$ & $110.53 \pm 13.57$ & Distribution criteria$^{a}$ & Distribution criteria$^{a}$ \\
DBP, mmHg & $75.28 \pm 12.83$ & $74.21 \pm 10.17$ & Distribution criteria$^{a}$ & Distribution criteria$^{a}$ \\
Sampling rate, Hz & 500 & 250 & -- & -- \\
Average signal length, s & 30 & 60 & -- & -- \\
Hypertension ratio (SBP, DBP) & 27.83\%, 12.37\% & 2.14\%, 8.36\% & $\geq 21$ in each of 4 BP groups & Reference BP distribution$^{a}$ \\
Measurement setting & In-office & Home-based long-term & Seated-rest/BP-change tests & Seated-rest paired protocol \\
Reference BP & Auscultatory device & Oscillometric device & Manual auscultation & Two-observer auscultation \\
\bottomrule
\end{tabularx}
\parbox{0.98\textwidth}{\scriptsize

\textit{Notes:}ISO BP-coverage thresholds are $\geq 5$/$20$/$5\%$ at SBP $\leq 100$/$\geq 140$/$\geq 160$ and DBP $\leq 60$/$\geq 85$/$\geq 100$ mmHg. }
\end{table*}

\subsubsection{Signal Conditioning and Feature Extraction}
For both datasets, PPG signals were processed using a unified signal-conditioning and feature-extraction pipeline before model training. Raw PPG signals were first filtered by a Butterworth band-pass filter with a passband of 0.25--10 Hz to suppress baseline drift and high-frequency noise. Baseline correction was then performed using the adaptive iteratively reweighted penalized least squares algorithm \cite{ref17}. The preprocessed PPG signals were downsampled to 125 Hz and normalized using min--max scaling.

Following Cisnal et al. \cite{ref18}, handcrafted PPG features were first extracted from each valid pulse beat within a preprocessed PPG segment and then averaged across all valid beats to obtain one segment-level feature vector. The resulting 476 handcrafted PPG features comprised 404 time-domain features, 58 frequency-domain features, and 14 statistical features. The features were normalized feature-wise using min--max scaling, with the minimum and maximum values estimated exclusively from the training set and then applied unchanged to the validation and test sets to prevent data leakage. No feature selection or dimensionality reduction was performed.

\subsection{Signal-Context Gated Fusion Architecture}
\begin{figure*}[!t]
  \centering
  \includegraphics[width=0.60\textwidth]{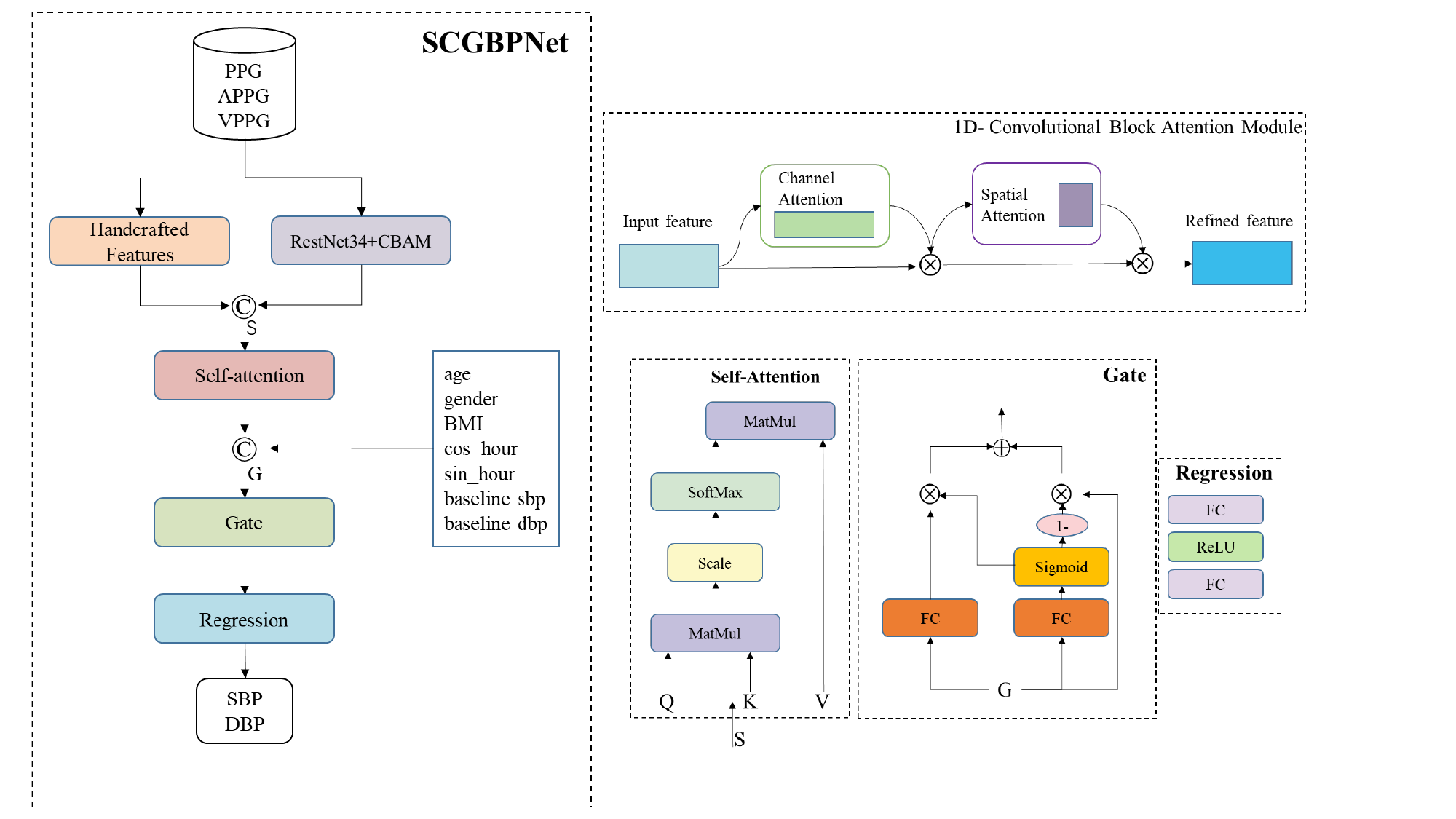}
  \caption{The proposed SCGBPNet architecture.}
  \label{fig:architecture}
\end{figure*}

The proposed Signal-Context Gated Fusion Network (SCGBPNet) is designed to integrate heterogeneous information from waveform signals, handcrafted features, and subject-specific context, including demographics, acquisition timing, and baseline BP. As shown in Fig.~\ref{fig:architecture}, SCGBPNet first fuses handcrafted and deep PPG representations at the signal level using self-attention, and then combines subject-specific context through a gated fusion module.

\subsubsection{Signal-level Fusion}
A 1-D ResNet34 backbone with Convolutional Block Attention Module (CBAM) was used to extract deep waveform representations from PPG, its first-order derivative (VPG), and its second-order derivative (APG). The deep waveform representations and the handcrafted PPG features were then stacked as a two-token representation and fused using a self-attention module. The attended output was then flattened and projected to a 256-dimensional signal-level representation.

\subsubsection{Subject-Context Gated Fusion}
Inspired by the gating mechanism of gated recurrent units \cite{ref19}, the fused signal representation was integrated with subject-specific context through a lightweight gated fusion module. Let $\mathbf{x}$ denote the concatenation of the self-attention-based signal features, non-baseline subject-specific context, and the baseline BP vector. The module generated dimension-wise gate weights $\mathbf{g}$ and a candidate representation $\mathbf{h}$:

\begin{equation}
  \mathbf{g} = \sigma\left(W_g \mathbf{x} + \mathbf{b}_g\right), \qquad
  \mathbf{h} = W_h \mathbf{x} + \mathbf{b}_h,
\end{equation}
\begin{equation}
  \mathbf{z} = \mathbf{g} \odot \mathbf{h} + (1-\mathbf{g}) \odot \mathbf{x}.
\end{equation}
Here, $\sigma$ denotes the sigmoid function and $\odot$ denotes element-wise multiplication. The final representation $\mathbf{z}$ was obtained as a gated residual mixture of the candidate embedding and the original input, allowing subject-specific context to be incorporated while preserving signal-level information.

\subsubsection{Regression Head}
The final fused representation was fed into a regression head to estimate SBP and DBP. To explicitly incorporate individual calibration information, baseline BP was concatenated with the fused representation before the final fully connected regression layer. Under baseline-only calibration, this baseline input remained fixed for each subject and was defined by the first valid reference BP measurement. Under periodic recalibration, it was updated using the first reference measurement within each predefined calibration window.

\subsection{Experimental Design}
For within-dataset evaluation, five-fold cross-validation was performed on both datasets using subject-wise partitioning to prevent data leakage. The model was optimized with the MAE loss. For cross-dataset evaluation, five models were trained on Aurora-BP. 10\% of the participants in the private dataset was held out as a validation set for selecting the best checkpoint, while the remaining 90\% formed an independent test set. The baseline BP input was defined as the first valid cuff-based reading in the private dataset and as the dataset-provided average of repeated readings in Aurora-BP. Under periodic recalibration, this input was updated using the first reference measurement collected within each predefined calibration window.

Under the Aurora-BP within-dataset setting, we performed architecture and input ablation studies. ResNet34 served as the architectural baseline, with compact self-attention (SA), CBAM, and the gated fusion module added progressively. For input ablation, we evaluated PPG-only, handcrafted-feature-only, static-information-only, and baseline-BP-only variants, as well as a model without baseline BP. These variants used the same architecture and training protocol as the full model. Parameter counts were reported to assess the accuracy-complexity trade-off.

For the private dataset, long-term stability was further analyzed under baseline-only calibration by evaluating week-to-week variation and daily performance decay analysis. In addition, six recalibration intervals were investigated to quantify the effect of recalibration frequency: daily, 3-day, 7-day, 14-day, 30-day, and baseline-only calibration.

\subsection{Evaluation Metrics}

Model performance was evaluated using mean absolute error (MAE), mean error (ME), the standard deviation (SD) of the estimation error, and the Pearson correlation coefficient between estimated and reference BP values. In addition to overall accuracy, two long-term stability metrics were introduced: weekly variation index (WVI) and daily performance-decay slope.

For WVI, weekly MAE was first computed by aggregating all valid samples within study week $k$ across subjects:
\begin{equation}
  \mathrm{MAE}_{k} = \frac{1}{N_k}\sum_{i\in \mathcal{S}_k}\left| y_i - \hat{y}_i \right|,
\end{equation}
where $\mathcal{S}_k$ denotes the set of valid samples in week $k$ and $N_k$ is the corresponding number of samples. Following ASEP-Net \cite{ref20}, WVI was calculated as the dispersion of weekly MAE values:
\begin{equation}
  \mathrm{WVI} = \sqrt{\frac{1}{K}\sum_{k=1}^{K}\left(\mathrm{MAE}_{k} - \overline{\mathrm{MAE}}\right)^2 }.
\end{equation}
Here, $K$ denotes the number of study weeks and $\overline{\mathrm{MAE}}$ denotes the mean weekly MAE.

For daily performance decay analysis, daily MAE was first obtained by aggregating all valid samples collected on the same study day. Daily performance decay after baseline calibration was then estimated by fitting daily MAE against days since baseline calibration:
\begin{equation}
  \mathrm{MAE}_{d} = \beta_0 + \beta_1 d + \epsilon_d,
\end{equation}
where $\beta_1$ represents the daily performance-decay slope in mmHg/day.

For the participant-level analysis, a total of 108 participants with measurements on at least 10 study days over a period of at least 21 days were included. Individual MAEs were calculated using all post-baseline measurements, and participant-specific daily decay slopes were estimated by regressing daily MAE per subject against days since baseline. 

\subsection{Implementation Details}
All experiments were implemented in PyTorch and executed on an NVIDIA Tesla V100 PCIe GPU with 32 GB memory. Models were trained for up to 200 epochs using the Adam optimizer with a batch size of 128. The learning rate and weight decay were set to $10^{-3}$. To mitigate overfitting, early stopping with a patience of 20 epochs was applied based on validation performance. In each fold, the checkpoint with the lowest validation MAE was selected for testing.

\begin{table*}[!t]
\centering
\caption{Performance Comparison, Cross-Dataset Evaluation, and Ablation Results for SCGBPNet}
\label{tab:regression_comparison}
\renewcommand{\arraystretch}{1.0}
\footnotesize
\setlength{\tabcolsep}{2pt}
\begin{tabular}{llcccccc}
\toprule
\multirow{2}{*}{Dataset setting} & \multirow{2}{*}{Model} & \multicolumn{2}{c}{SBP} & \multicolumn{2}{c}{DBP} & \multirow{2}{*}{Params (M)} & \multirow{2}{*}{\makecell{Computational\\load (GFLOPs)}} \\
\cmidrule(lr){3-4}\cmidrule(lr){5-6}
 & & MAE (mmHg) & ME$\pm$SD (mmHg) & MAE (mmHg) & ME$\pm$SD (mmHg) & & \\
\midrule
\multicolumn{8}{l}{\textit{Baseline model comparison}} \\
Aurora-BP CV & Ridge \cite{ref6} & -- & $0.42 \pm 10.10$ & -- & $0.55 \pm 7.66$ & -- & -- \\
Aurora-BP CV & Decision Tree & $8.48^{*}$ & $-0.14 \pm 12.42$ & $5.78^{*}$ & $0.19 \pm 8.04$ & -- & -- \\
Aurora-BP CV & Gradient Boost \cite{ref18} & 8.49 & $0.63 \pm 11.73$ & 5.03 & $0.01 \pm 6.58$ & -- & -- \\
Aurora-BP CV & Adaboost & $7.65^{*}$ & $-0.47 \pm 10.78$ & $5.30^{*}$ & $0.47 \pm 7.20$ & -- & -- \\
Aurora-BP CV & ResNet50 \cite{ref21} & 7.47 & $0.85 \pm 11.05$ & $5.13^{*}$ & $-0.18 \pm 7.12$ & -- & -- \\
Aurora-BP CV & VGGNet16 \cite{ref22} & 7.83 & $1.67 \pm 12.02$ & $5.25^{*}$ & $-0.74 \pm 7.50$ & -- & -- \\
Aurora-BP CV & HGCTNet \cite{ref15} & 7.61 & $1.24 \pm 11.52$ & 5.21 & $-0.90 \pm 7.43$ & -- & -- \\
Aurora-BP CV & SCGBPNet (\textit{Ours}) & \textbf{7.08} & $0.96 \pm \textbf{10.29}$ & \textbf{4.90} & $-0.46 \pm \textbf{6.78}$ & 8.277 & 0.226 \\
\midrule
\multicolumn{8}{l}{\textit{Architecture ablation}} \\
Aurora-BP CV & ResNet34 & 7.47 & $0.85 \pm 11.05$ & 5.13 & $-0.18 \pm 7.12$ & 7.294 & 0.224 \\
Aurora-BP CV & ResNet34+SA & 7.24 & $0.93 \pm 10.56$ & 4.99 & $-0.21 \pm 6.91$ & 8.020 & 0.225 \\
Aurora-BP CV & ResNet34+CBAM+SA & 7.14 & $0.48 \pm 10.37$ & 4.94 & $0.14 \pm 6.80$ & 8.177 & 0.226 \\
Aurora-BP CV & Full SCGBPNet & \textbf{7.08} & $0.96 \pm \textbf{10.29}$ & \textbf{4.90} & $-0.46 \pm \textbf{6.78}$ & 8.277 & 0.226 \\
\midrule
\multicolumn{8}{l}{\textit{Input ablation}} \\
Aurora-BP CV & w/o baseline BP & 12.38 & $0.79 \pm 15.89$ & 8.66 & $0.28 \pm 10.96$ & 8.277 & 0.226 \\
Aurora-BP CV & PPG only & 13.91 & $2.01 \pm 17.78$ & 9.06 & $0.93 \pm 11.52$ & 8.277 & 0.226 \\
Aurora-BP CV & Handcrafted features only & 13.06 & $1.00 \pm 16.85$ & 8.95 & $0.95 \pm 11.47$ & 8.277 & 0.226 \\
Aurora-BP CV & Demographics/BMI only & 13.73 & $1.90 \pm 17.89$ & 9.05 & $0.53 \pm 11.65$ & 8.277 & 0.226 \\
Aurora-BP CV & Baseline BP only & 7.92 & $1.77 \pm 12.21$ & 5.26 & $-0.85 \pm 7.51$ & 8.277 & 0.226 \\
\midrule
\multicolumn{8}{l}{\textit{Within- and cross-dataset evaluation}} \\
Smartwatch CV & SCGBPNet (\textit{Ours}) & 6.44 & $-0.12 \pm 8.28$ & 4.89 & $0.02 \pm 6.24$ & 8.277 & 0.226 \\
Aurora-BP$\rightarrow$Smartwatch & SCGBPNet (\textit{Ours}) & 8.98 & $-5.32 \pm 10.80$ & 6.47 & $-3.16 \pm 7.71$ & 8.277 & 0.226 \\
\bottomrule
\multicolumn{8}{l}{\footnotesize CV: five-fold subject-wise cross-validation.} \\
\multicolumn{8}{l}{\footnotesize $^{*}$Significant difference ($p<0.05$) in MAE compared with SCGBPNet under Mann-Whitney U test.} \\
\end{tabular}
\end{table*}

\section{Results}
\subsection{Overall Accuracy}
Table~\ref{tab:regression_comparison} summarizes the baseline model comparison, ablation study, within- and cross-dataset evaluation. In the Aurora-BP baseline comparison, SCGBPNet achieved the lowest MAEs for both SBP and DBP, reaching 7.08 and 4.90 mmHg, respectively. Under within-dataset evaluation, DBP accuracy was comparable between the private smartwatch and Aurora-BP datasets (4.89 vs. 4.90 mmHg), whereas SBP MAE was lower on the smartwatch dataset (6.44 vs. 7.08 mmHg). The corresponding correlation coefficients were 0.87 for SBP and 0.85 for DBP on the Aurora-BP dataset, and 0.79 for both SBP and DBP on the private smartwatch dataset.

Cross-dataset transfer from Aurora-BP to the smartwatch dataset resulted in higher SBP/DBP MAEs of 8.98/6.47 mmHg and MEs of $-5.32$/$-3.16$ mmHg, indicating systematic underestimation.

\subsection{Ablation Results}
The architecture ablation results in Table~\ref{tab:regression_comparison} demonstrate incremental gains from the proposed fusion components beyond the ResNet34 backbone. Compact self-attention reduced SBP/DBP MAE from 7.47/5.13 to 7.24/4.99 mmHg, while adding CBAM and the subject-context gated fusion module further reduced the errors to 7.08/4.90 mmHg with limited parameter growth and computational overhead.

The input ablation results showed that removing baseline BP significantly increased the SBP/DBP MAEs from 7.08/4.90 to 12.38/8.66 mmHg. Baseline BP alone yielded MAEs of 7.92/5.26 mmHg, while the other single-input models showed substantially higher errors. This comparison shows that baseline BP was a major contributor to prediction accuracy, while the additional inputs provided a further improvement.

\subsection{Long-Term Stability Analysis}

\begin{figure*}[!t]
  \centering
  \includegraphics[width=0.72\textwidth]{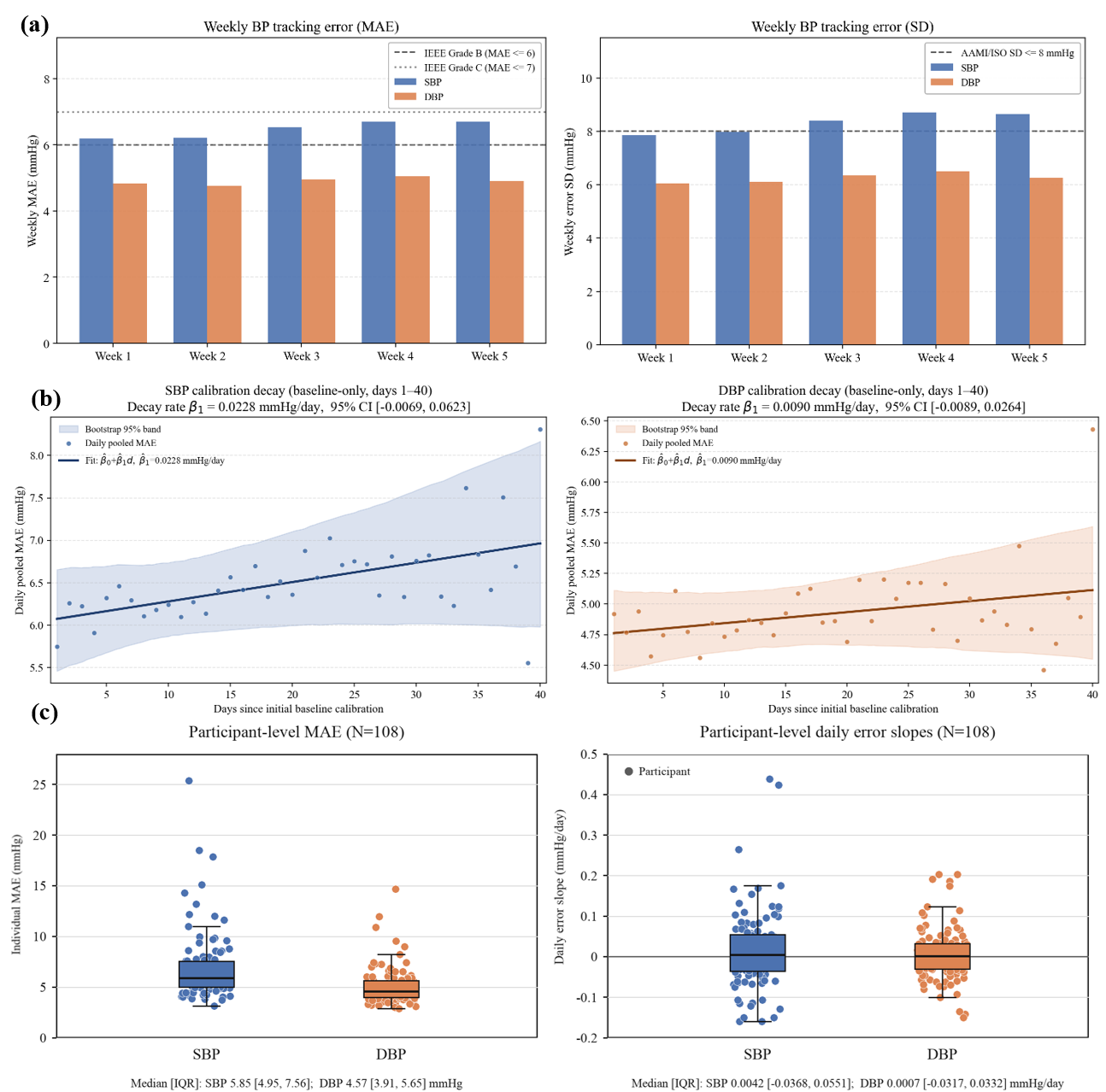}
  \caption{Long-term stability analysis under baseline-only calibration on the private smartwatch dataset. (a) Weekly MAE and weekly error SD for SBP and DBP; (b) pooled daily performance-decay curves with bootstrap 95\% confidence bands; (c) participant-level MAE and daily decay slopes. Points represent participants, boxes show the median and IQR.}
  \label{fig:weekly_daily_stability}
\end{figure*}

\subsubsection{Week-to-Week Tracking Stability}
As shown in Fig.~\ref{fig:weekly_daily_stability}(a), SBP weekly MAE increased from 6.19 mmHg in Week 1 to 6.71 mmHg in Weeks 4--5, whereas DBP weekly MAE remained within 4.77--5.06 mmHg. Relative to the threshold lines, DBP weekly MAE remained within IEEE Grade B (<6 mmHg) and its weekly error SD stayed below the AAMI/ISO threshold of 8 mmHg, while SBP weekly MAE remained within Grade C and its weekly error SD exceeded 8 mmHg from Week 3 onward. The resulting WVI values were 0.2295 mmHg for SBP and 0.1020 mmHg for DBP, indicating smaller week-level fluctuation for DBP.

\subsubsection{Daily Performance Decay and Recalibration Frequency}
As shown in Fig.~\ref{fig:weekly_daily_stability}(b), the linear trend analysis yielded estimated slopes of 0.0228 mmHg/day for SBP and 0.0090 mmHg/day for DBP, suggesting only a slight upward trend in estimation error over days. Because the 95\% confidence intervals for both slopes included zero, the null hypothesis of no significant performance decay could not be rejected within the predefined observation window. These results indicate that baseline-only calibration did not exhibit statistically confirmed time-dependent performance degradation. 

Fig.~\ref{fig:weekly_daily_stability}(c) summarizes the participant-level MAE and daily decay slopes for the 108 participants. The median MAEs were 5.85 mmHg (IQR: 4.95--7.56 mmHg) for SBP and 4.57 mmHg (IQR: 3.91--5.65 mmHg) for DBP, indicating substantial interparticipant variability. The corresponding median daily slopes were 0.0042 mmHg/day (IQR: $-0.0368$ to 0.0551 mmHg/day) and 0.0007 mmHg/day (IQR: $-0.0317$ to 0.0332 mmHg/day), respectively. Except for a few outliers, most participant-level slopes were concentrated within $[-0.2, 0.2]$ mmHg/day, indicating overall good stability for the test population.

For the recalibration schemes in Table~\ref{tab:recalibration}, DBP performance varied only modestly across intervals, suggesting that frequent recalibration was unnecessary for DBP. For SBP, weekly recalibration was approximately required to approach the AAMI/ISO SD criterion; IEEE Grade C was maintained for up to one month without recalibration, whereas Grade B required daily recalibration. However, the absolute benefit of daily recalibration was limited during the observation window, with MAE reductions of only 0.51 mmHg for SBP and 0.35 mmHg for DBP compared with baseline-only calibration.

\subsubsection{Sensitivity to Individual Baseline Deviation}
\begin{figure}[!t]
  \centering
  \includegraphics[width=1.00\columnwidth]{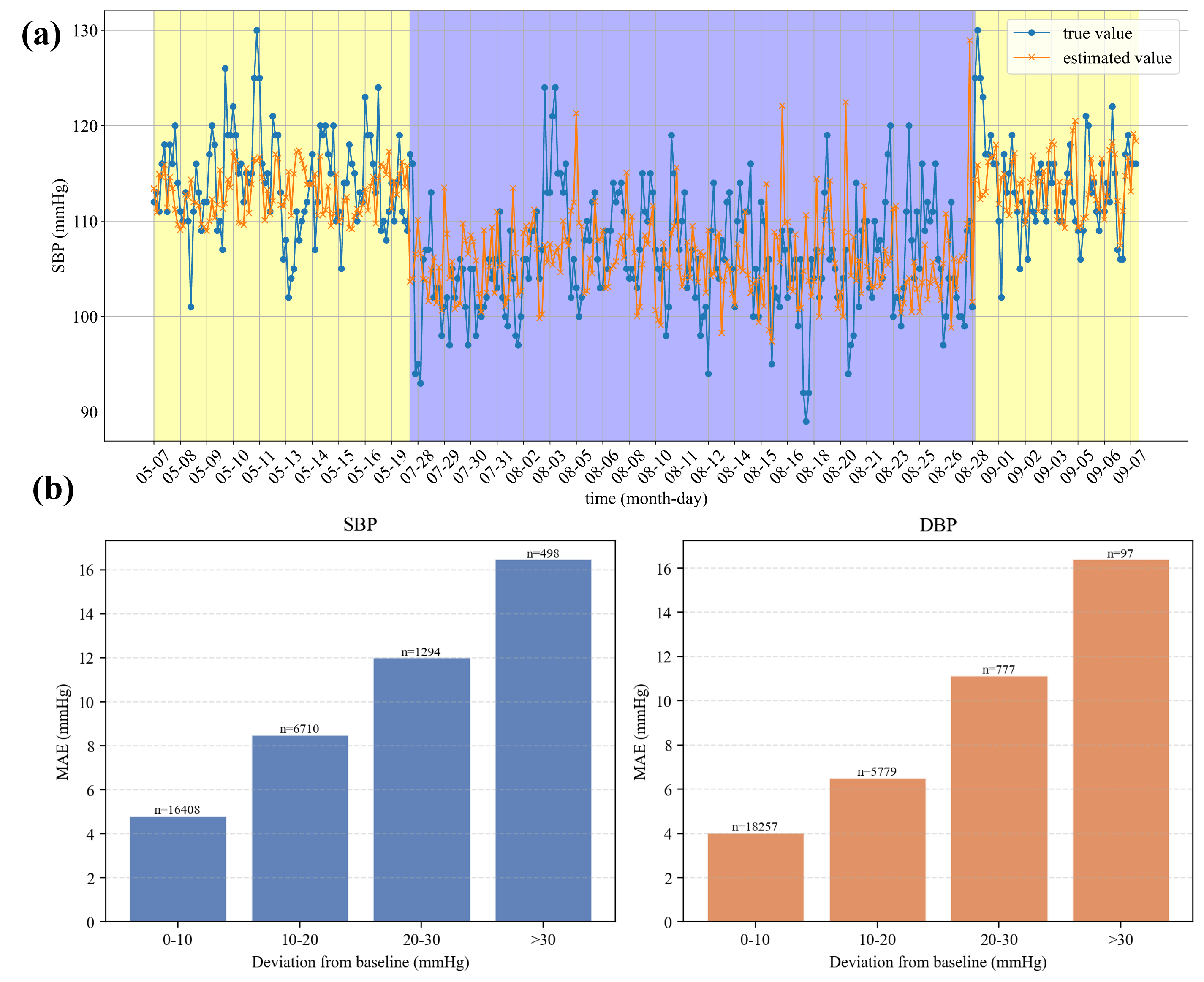}
  \caption{Individual BP tracking and baseline-deviation sensitivity for the private dataset. (a) The reference and estimated SBP trajectories; (b) MAE at different deviation levels from individual baseline BP for SBP and DBP.}
  \label{fig:private_deviation_mae}
\end{figure}

Fig.~\ref{fig:private_deviation_mae} summarizes individual BP tracking and baseline-deviation sensitivity in the private smartwatch dataset. As shown in Fig.~\ref{fig:private_deviation_mae}(a), the estimated SBP trajectory broadly followed the reference trend but was smoother and did not fully capture abrupt BP changes. Fig.~\ref{fig:private_deviation_mae}(b) further stratifies samples by their absolute deviation from each subject's baseline BP. Among the 24910 samples, 16408 were within 10 mmHg of the individual baseline. For both SBP and DBP, MAE increased progressively with larger baseline deviations, and only the within-10-mmHg subgroup satisfied the predefined accuracy requirement. This predominance of near-baseline samples likely contributed to the favorable overall accuracy. These findings indicate that baseline-calibrated estimation was most reliable near the calibration state but became less accurate as BP shifted farther from the individual baseline.

\begin{table}[!htbp]
\centering
\caption{Model Performance at Different Recalibration Intervals on the Private Smartwatch Dataset}
\label{tab:recalibration}
\renewcommand{\arraystretch}{1.1}
\scriptsize
\setlength{\tabcolsep}{3pt}
\begin{tabular}{lcccc}
\toprule
\multirow{2}{*}{Interval} & \multicolumn{2}{c}{SBP} & \multicolumn{2}{c}{DBP} \\
\cmidrule(lr){2-3}\cmidrule(lr){4-5}
 & MAE (mmHg) & ME$\pm$SD (mmHg) & MAE (mmHg) & ME$\pm$SD (mmHg)\\
\midrule
1 Day & 5.93 & $1.97\pm7.43$ & 4.54 & $1.55\pm5.59$ \\
3 Days & 6.26 & $1.98\pm7.83$ & 4.83 & $1.64\pm5.97$ \\
7 Days & $6.38^{*}$ & $1.59\pm8.07$ & $4.97^{*}$ & $1.32\pm6.07$ \\
14 Days & $6.40^{*}$ & $1.10\pm8.14$ & $4.91^{*}$ & $0.95\pm6.18$ \\
30 Days & $6.42^{*}$ & $0.11\pm8.24$ & $4.87^{*}$ & $0.20\pm6.21$ \\
Baseline & $6.44^{*}$ & $-0.12\pm8.28$ & $4.89^{*}$ & $0.02\pm6.24$ \\
\bottomrule
\multicolumn{5}{@{}p{\dimexpr\columnwidth-2\tabcolsep\relax}@{}}{\footnotesize $^{*}$Significant difference ($p<0.05$) compared with daily calibration using a paired $t$-test.}
\end{tabular}
\end{table}

\section{Discussion}

This study proposed SCGBPNet, an efficient signal-context gated fusion framework for cuffless BP estimation from heterogeneous BP-related inputs. The ablation results indicate that the proposed fusion design effectively combines all inputs, improving accuracy with limited computational overhead. With further optimization of the lightweight backbone, model compression, and quantization-aware evaluation, this architecture may be suitable for future edge-device deployment. 

Comprehensive validation through subject-wise cross-validation within the Aurora-BP and private smartwatch datasets, as well as cross-dataset transfer from Aurora-BP to the smartwatch dataset, showed that accurate BP estimation remains challenging across heterogeneous populations and monitoring settings. The within-dataset performance gap, together with the marked degradation observed in cross-dataset evaluation, may be caused by the different characteristics of the two datasets, including population characteristics, BP distributions, acquisition protocols, and wearable hardware. Consequently, current cuffless BP models remain highly dataset- and device-dependent, posing substantial challenges for deployment in large real-world cohorts.

Overall, the results demonstrated acceptable long-term stability at both the aggregate and individual-participant levels over a follow-up period of more than one month, despite a small number of outliers. Compared with baseline-only calibration, daily recalibration only brought limited improvements, the potential accuracy benefit of frequent recalibration should be weighed against user burden and continued dependence on cuff measurements. 

Compared with previous smartwatch-based stability studies, the Samsung smartwatch study evaluated real-world measurement stability based on estimated BP drift and recommended a 14-day recalibration interval \cite{ref16}. However, it was not designed as a standard accuracy-validation trial and did not collect cuff-based reference BP on each day of the observation window. In contrast, the present study uses paired cuff reference and smartwatch PPG measurements to quantify daily performance decay and recalibration effects, providing more direct evidence for recalibration assessment. Another recent study, ASEP-Net, reported WVI values of 0.68/0.92 mmHg for SBP/DBP on a one-month smartwatch dataset \cite{ref20}. In comparison, our method achieved lower WVI values of 0.2295/0.1020 mmHg, together with low daily decay. Future work should validate these findings in larger and more diverse populations.

In addition, our findings indicate that it remains challenging when BP deviates substantially from the individual baseline. Therefore, future studies of cuffless BP models should report stratified performance across BP ranges, baseline-deviation levels, and population subgroups.

\section{Conclusion and Future Work}
This study presented SCGBPNet, an efficient signal-context gated fusion architecture for smartwatch-based cuffless BP monitoring. The discrepancy between within-dataset validation performance and cross-dataset degradation indicates that acceptable accuracy can currently be achieved only under dataset-specific conditions with varied performance across experimental conditions, posing a major challenge to real-world deployment. These findings suggest that a more systematic understanding of how population characteristics, BP distributions, and individual BP excursions affect model performance is needed to inform standardized evaluation protocols for deployable cuffless BP systems. 

\section*{Acknowledgment}
This work was supported by the Young Scientists Fund from the National Natural Science Foundation of China (NSFC) [62301333], the National Natural Science Foundation of China (NSFC) [62373259], and the Special Project for Major Science and Technology Initiatives of the Tibet Science and Technology Department [XZ202201ZD0001G01].

\end{document}